\documentclass[twocolumn,preprintnumbers,amsmath,amssymb,aps,prl]{revtex4-2}

\usepackage{graphicx}
\usepackage{dcolumn}
\usepackage{bm}
\usepackage[hidelinks]{hyperref}
\usepackage{setspace}
\usepackage{subcaption}
\begin{document}

\title{Theory for the power threshold of the L--H transition in tokamaks}

\author{B. De Lucca}
\email{brenno.delucca@epfl.ch}
\affiliation{\'Ecole Polytechnique F\'ed\'erale de Lausanne (EPFL), Swiss Plasma Center (SPC), CH-1015 Lausanne, Switzerland}

\author{P. Ricci}
\affiliation{\'Ecole Polytechnique F\'ed\'erale de Lausanne (EPFL), Swiss Plasma Center (SPC), CH-1015 Lausanne, Switzerland}

\author{B. Labit}
\affiliation{\'Ecole Polytechnique F\'ed\'erale de Lausanne (EPFL), Swiss Plasma Center (SPC), CH-1015 Lausanne, Switzerland}

\author{D. Mancini}
\affiliation{\'Ecole Polytechnique F\'ed\'erale de Lausanne (EPFL), Swiss Plasma Center (SPC), CH-1015 Lausanne, Switzerland}

\author{L. N. Stenger}
\affiliation{\'Ecole Polytechnique F\'ed\'erale de Lausanne (EPFL), Swiss Plasma Center (SPC), CH-1015 Lausanne, Switzerland}

\author{Z. Tecchiolli}
\affiliation{\'Ecole Polytechnique F\'ed\'erale de Lausanne (EPFL), Swiss Plasma Center (SPC), CH-1015 Lausanne, Switzerland}

\date{\today}

\begin{abstract}
    Flux-driven two-fluid simulations in a diverted tokamak exhibit a confinement transition when a critical heating threshold is exceeded, as in experimental observations of the L--H transition. The simulations show that electromagnetic drift-wave turbulence spontaneously generates a sheared $\bm{E}\times \bm{B}$ flow responsible for transport suppression in the edge. The toroidal-field asymmetry of the transition, here reproduced in simulations for the first time, arises from time-reversal symmetry breaking by finite collisionality, as demonstrated by a quasilinear calculation of the turbulent momentum flux. Scaling laws are analytically derived for the L--H power threshold in the high- and low-density branches, for the density minimum, and for the minimum power, all matching or surpassing the accuracy of existing empirical scalings.
\end{abstract}

\maketitle

Above a threshold input power, turbulence in the edge region of tokamaks undergoes a rapid suppression, leading to a factor $\sim 2$ increase in the energy confinement time. Known as the L--H transition~\cite{Wagner1982}, this mechanism is considered essential to achieve burning plasma conditions in future devices, including ITER~\cite{Doyle_2007} and SPARC~\cite{Creely_2020}. Despite numerous theoretical proposals~\cite{Diamond1994, Pogutse1997,Rogers1998,Guzdar2001,Kim2003, Fundamenski2012, Ryter2014, Bilato_2020, Eich2021}, the physics underlying the L--H transition remains debated, as no existing model simultaneously accounts for several well-established experimental observations. These include the non-monotonic dependence of the threshold power, $P_{\mathrm{LH}}$, on plasma density~\cite{Fielding1996} and the observation that the threshold power is generally lower when the ion $\nabla B$ drift points from the core region toward the \textit{X}-point in a diverted geometry (\textit{favourable} configuration)~\cite{ASDEX1989, Takizuka2004}. For densities above a critical value, called the density minimum, $P_{\mathrm{LH}}$ is well-described by the empirical scaling law~\cite{Martin2008}:
\begin{equation}
P_{\mathrm{LH}}^{\rm ITPA} = 0.0488\, \overline{n}_{e20}^{0.717}  B^{0.803} S^{0.941}\,[\text{MW}],
\label{eq:ITPA}
\end{equation}
with $\overline{n}_{e20}$ the line-averaged density in $10^{20} \, \mathrm{m}^{-3}$, $B$ the magnetic field strength in T, and $S$ the plasma surface area in $\mathrm{m}^2$.
Though edge-transport bifurcations have been observed several times in global flux-driven turbulence simulations~\cite{Rasmussen2015, Li2017, Chang2017, Giacomin2020, DifPradalier2022, Zholobenko2026}, their relevance to the experimental L--H transition observations is subject to debate, and predictive physics-based capabilities are lacking.

Here, we present power-ramp simulations, starting from L--mode turbulence, carried out with the global flux-driven code \texttt{GBS}~\cite{Ricci2012, Giacomin2022}, which solves the drift-reduced Braginskii equations. These simulations exhibit a confinement transition occurring at lower power in the favourable toroidal-field configuration. Based on these results, we derive a theory of the L--H transition, providing, to our knowledge, the first unified analytic prediction of the power threshold, its toroidal-field asymmetry, and the density minimum. 

The physical model solved by \texttt{GBS} is reported in detail in~\cite{Giacomin2022}. Using a fluid model is justified by the large edge collisionality in existing and future tokamaks: for typical L--mode parameters in TCV~\cite{Hofmann_1994, Theiler_2026}, the ratio of electron mean-free-path to connection length is small in the edge, $\lambda_e/L_\parallel \lesssim 0.1$, with comparable values in AUG~\cite{Rogers1998}, and similar values expected in ITER and SPARC. The simulation parameters at the separatrix are $R/\rho_s = 500$, $\nu = 5 \times 10^{-2} c_s/R$, $a/R \simeq 0.3$, $Z = 1$, $m_i/m_e = 800$, and $\beta = 10^{-4}$, where $R$ is the tokamak major radius, $\rho_s = c_s/\Omega_i$ is the sound Larmor radius, with $c_s = \sqrt{T_e/m_i}$ the ion sound speed and $\Omega_i = Z e B/m_i$ the cyclotron frequency, $\nu = e^2 n/(m_i \sigma_\|)$ is the collisional frequency, $a$ is the minor radius, and $\beta = 2 \mu_0 p_e/B^2$ is the electron plasma beta. The magnetic field $\bm{B} = \sigma_T R|B_\varphi|\nabla\varphi + \sigma_p\nabla\varphi \times \nabla\psi$ is that of a diverted lower-single-null tokamak, with edge safety factor $q_{95} \simeq 4$, magnetic shear $\hat s_{95} \simeq 2$, and previously used in Ref.~\cite{Giacomin2020}. The plasma current direction is encoded in $\sigma_p \equiv \mathrm{sign}(\bm{I}_p \cdot \nabla \varphi)$, defined such that $\psi$ is always an increasing function of the volume enclosed by the flux-surface. The favourable configuration corresponds to $\sigma_T \equiv \mathrm{sign}(\bm{B} \cdot \nabla \varphi) =-1$. The reduced $m_i/m_e$ used for numerical stability does not affect the results, which are governed by resistive dynamics evolving on timescales $\partial_t \ll (m_i/m_e)\nu$.
\begin{figure}[h]
    \centering
    \includegraphics[width=0.98\columnwidth]{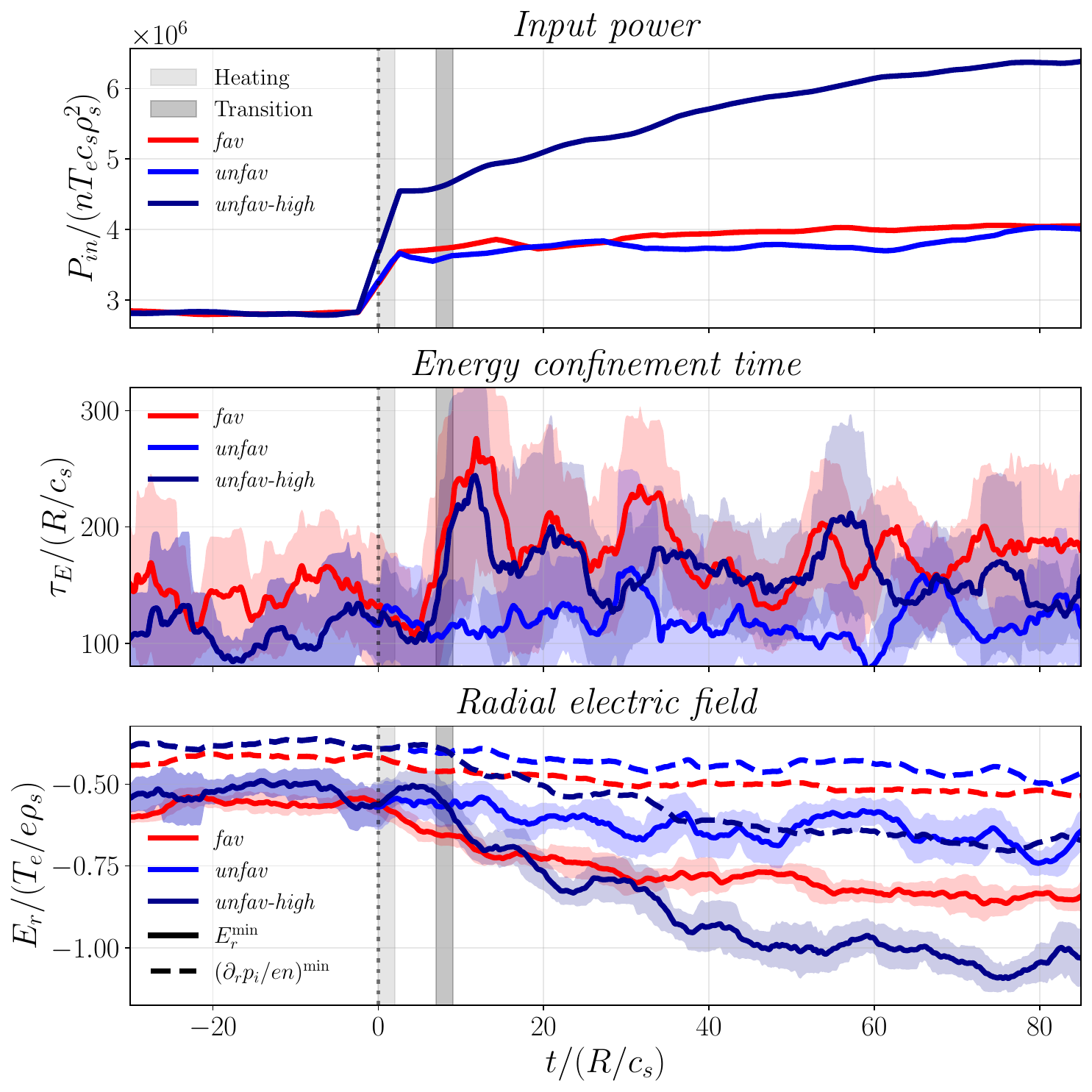}
    \caption{Time traces of injected power $P_{\rm in}$ (top), energy confinement time $\tau_E$ (middle), minima of radial electric field $E_r^{\rm min}$ and ion pressure gradient $(\partial_r p_i/en)^{\rm min}$ (bottom), for simulations in the favourable magnetic field direction (\textit{fav}, red), unfavourable (\textit{unfav}, blue) and unfavourable at higher power (\textit{unfav-high}, dark-blue).}
    \label{fig:time_traces}
\end{figure}

\begin{figure}[htbp]
    \centering
    \includegraphics[width=\columnwidth]{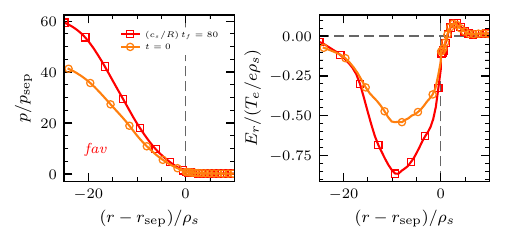} \\
        \includegraphics[width=\columnwidth]{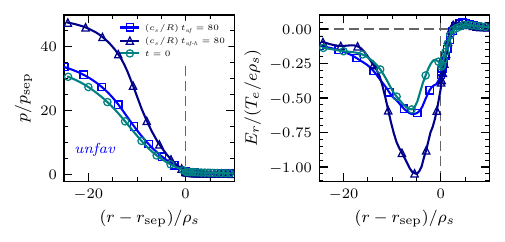}  
    \caption{Radial profiles of flux-surface-averaged total pressure $p$ (normalised to the separatrix value $p_{\rm sep}$) and radial electric field $E_r$, at $t=0$ and at $t=80 \,R/c_s$. Favourable (top) and unfavourable (bottom), with the latter shown at both matched input power (blue squares), and higher power (dark-blue triangles).}
    \label{fig:kinetic_profiles}
\end{figure}

Starting from L--mode simulations in quasi-steady-state, the input power is suddenly increased by a factor $\simeq 1.4$ at $t=0$ by raising the core heating sources (equal for ions and electrons) at constant density source, as shown in Fig.~\ref{fig:time_traces} (normalisations taken at the separatrix). In the favourable configuration, $\sim 10 \, R/c_s$ after additional heating is injected, a fast transport bifurcation occurs on a timescale of $\sim 5 \, R/c_s$, with $\tau_E$ increasing by a factor $\sim 2$. After an initial overshoot, the confinement time stabilises at a factor $\sim 1.5$ larger than its initial value. The flux-surface-averaged pressure and radial electric field profiles are shown in Fig.~\ref{fig:kinetic_profiles} as a function of the distance from the separatrix, $r-r_{\rm sep}$. During the transition, the radial electric-field well deepens, suppressing turbulent transport and steepening the edge pressure profile (Figs.~\ref{fig:time_traces} and~\ref{fig:kinetic_profiles}). Conversely, for the same input power, the unfavourable simulation shows neither improved confinement nor significant changes in the electric field. However, repeating the ramp at roughly twice the L--mode input power, the unfavourable case also transitions to a high confinement mode, with a similar deepening of the electric-field well as in the favourable case.
Edge turbulence prior to the transition arises from a combination of resistive ballooning and electron drift-wave (DW) modes. Removing the curvature drive from the simulations substantially suppresses turbulence, confirming the role of interchange modes, and spatio-temporal Fourier analysis reveals DW signatures: $\omega \simeq \omega_* $ and  $ e\widetilde{\phi}/\langle T_e \rangle \sim \widetilde{n}/\langle n \rangle \sim \widetilde{p}_e /\langle p_e \rangle$ (fluctuations are denoted by $\widetilde{f} = f - \langle f \rangle$, with $\langle f \rangle$ the time- and flux-surface average). 

A transport barrier only forms in the simulations when the edge plasma--$\beta$ exceeds a critical value, consistent with previous flux-tube results~\cite{Rogers1998}: the same power ramps at $\beta = 0$ produce no transition. Two electromagnetic mechanisms drive the suppression: linear DW stabilisation via enhanced electron adiabatic response~\cite{Rogers2005, Mosetto2012}, and nonlinear $\bm{E} \times \bm{B}$ shear generation by finite--$\beta$ DWs~\cite{Guzdar2001}. In fact, artificially removing the flux-surface-averaged component of the $\bm{E} \times \bm{B}$ drift velocity $\bm{v}_E$ ($\bm{v}_E \cdot \nabla \to (\bm{v}_E - \langle \bm{v}_E \rangle) \cdot \nabla$) immediately restores high transport levels, confirming its role in barrier formation and consistent with both experiments and previous simulations~\cite{Groebner1990,Cziegler2017,Chang2017,Zholobenko2026}. The rise of $\bm{E} \times \bm{B}$ shear originates from turbulent generation of momentum flux, as revealed by analysing the radial electric field evolution equation. 
\begin{figure}
  \centering
  \includegraphics[width=0.49\columnwidth]{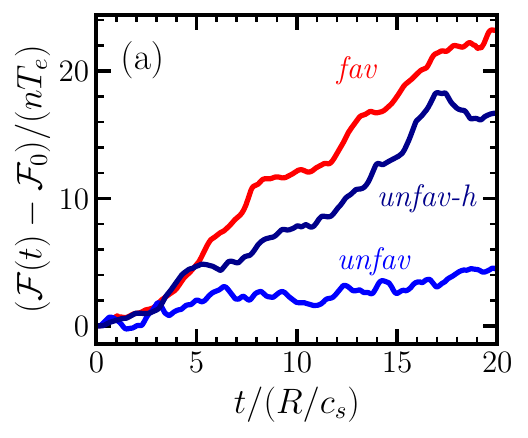}%
  \hfill
  \includegraphics[width=0.49\columnwidth]{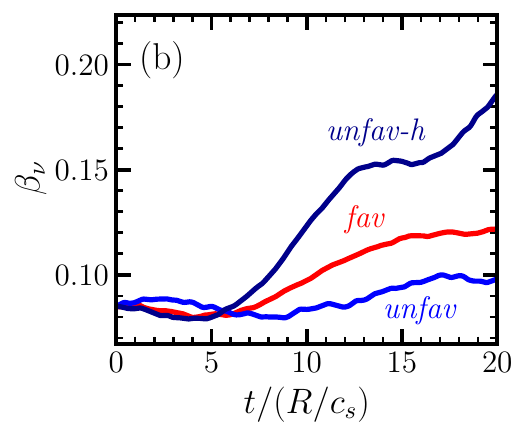}%
  \caption{(a) Turbulent momentum flux $\mathcal{F}(t) - \mathcal{F}_0$ following the onset of additional heating, with $\mathcal{F}_0 \equiv \mathcal{F}(t=0)$ subtracted to highlight the transition. (b) Time trace of $\beta_\nu$ at $r=r_{\rm min}$.}
\label{fig:fav_unfav_stresses_time}
\end{figure}
In \texttt{GBS}, the flux-surface-averaged radial electric field evolves as
$\partial_t \langle \mathcal{E}_r \rangle = \mathcal{F}$, where
$\mathcal{E}_r = (dr/d\psi)\, n m_i |\nabla \psi|^2 \partial_r \phi/B^2$ and $\mathcal{F} \simeq \langle
J_\parallel \delta \bm{B} \cdot \nabla \psi/B \rangle - \langle
\bm{v}_E \cdot \nabla \bm{\varpi} \cdot \nabla \psi \rangle$ encodes the nonlinear momentum flux (Maxwell and Reynolds stress). Here, $\bm{\varpi} = (n m_i/B^2)(\nabla_\perp \phi + \nabla_\perp p_i/e n)$ is the generalised vorticity and $\delta \bm{B} = \nabla \times (A_\parallel \bm{b})$ the perturbed magnetic field. Gyro-viscosity, parallel vorticity advection, radial diamagnetic current flux, and $\partial_t (\partial_r p_i/en)$ are neglected in the above expression for $\mathcal{F}$, as they yield a small contribution to $\partial_t \langle \mathcal{E}_r \rangle$ in the simulations. Figure~\ref{fig:fav_unfav_stresses_time}a shows the nonlinear momentum flux $\mathcal{F}$ at the position of the electric-field-well minimum, $r = r_{\rm min}$, as a function of time during the transition phase, $t \in (0, 20 \, R/c_s)$. The unfavourable configuration shows no change in the nonlinear drive of mean $\bm{E} \times \bm{B}$ flow unless a higher input power is used, as seen experimentally~\cite{Cziegler2017,Yan2023}. To explain these observations, we derive the conditions under which finite--$\beta$ DWs can drive the formation of a mean flow, generalising the work of~\cite{Rogers1998,Guzdar2001} and proving that the threshold condition is accessed, in a collisional plasma, at a lower input power in the favourable configuration.

Adopting a local straight-field-line coordinate system, centred at the $\psi = \psi_0$ flux-surface and given by $x = |q_0|(\psi - \psi_0)/(r_0 B)$, $y = \sigma_T \sigma_p r_0(q\theta-\varphi)/|q_0| + \sigma_T \hat{s} \Omega_E x t$ and $z = qR\theta$, the magnetic field is expressed as $\bm{B} = \sigma_T B\nabla x \times \nabla y$~\cite{Waelbroeck1991,Newton2010}. The coordinate $\theta$ is the ballooning angle, with $\theta = 0$ at the outer-midplane and $\theta=-\pi$ at the \textit{X}-point. The $\bm{E} \times \bm{B}$ shearing rate normalised to magnetic shear is given by $\Omega_E = -(\rho_s c_s |q| R/v_A) d_x^2\langle \phi \rangle/\hat{s} $, where $v_A = B/\sqrt{\mu_0 m_i \langle n \rangle}$ is the Alfvén velocity. Starting from the drift-reduced Braginskii model, assuming cold ions, weak curvature, and neglecting coupling to sound-waves ($k_\parallel c_s \ll \partial_t \sim \omega_* = k_y \rho_s c_s |d_x \log \langle n \rangle |$), we derive the system describing the DW linear instability, yielding 
\begin{align}
    &\frac{\partial n}{\partial t} + i\sigma_T \rho_s k_y \sqrt{\beta_A}\phi + \sigma_p  \frac{\partial( \rho_s^2 k_\perp^2 A_\parallel)}{\partial \theta} = 0, \label{eq:density_main_DW}\\
    &\frac{\partial (\rho_s^2 k_\perp^2 \phi)}{\partial t} - \sigma_p  \frac{\partial(\rho_s^2 k_\perp^2 A_\parallel) }{\partial \theta} = 0, \label{eq:vorticity_main_DW}\\
    &\frac{\partial }{\partial t}\left(1 + \frac{\rho_s^2 k_\perp^2}{\beta_\mu} \right)A_\parallel + i \sigma_T \rho_s k_y \sqrt{\beta_A}A_\parallel  \nonumber \\
    &\qquad - \sigma_p \frac{\partial(\phi -n)}{\partial \theta} + \frac{\rho_s^2 k_\perp^2 }{\beta_\nu}A_\parallel = 0, \label{eq:vpare_main_DW} 
\end{align}
where the perpendicular wavenumber, 
\begin{equation}\label{eq:kperp}
k_\perp^2 = k_y^2 \left[ \frac{1}{g} + \hat{s}^2\left(\theta +  \frac{r_0}{\hat s} \partial_x \theta - \sigma_T \Omega_E t\right)^2 g \right] ,
\end{equation}
acquires its characteristic time-dependence in a radially-varying background flow~\cite{Waelbroeck1991, Newton2010, Parra2011, Schekochihin2012},
with modes sheared along the field-lines by $\hat s$, and in time by $\bm{E} \times \bm{B}$ shear. 
  In Eqs.~(\ref{eq:density_main_DW}--\ref{eq:kperp}), we denote the radial metric element as $g(\theta, r_0) \equiv |\nabla x|^2$ and the fields are normalised as $e\phi/\langle T_e \rangle \to \phi $, $[\Omega_i v_A/(c_s^2 B)] A_\parallel \to A_\parallel $, $ n/\langle n \rangle \to n$, with time normalised to the Alfvén frequency, $\omega_A= v_A/(|q|R)$. Three main parameters regulate the dynamics in Eqs.~(\ref{eq:density_main_DW}--\ref{eq:vpare_main_DW}): $\beta_\nu = \beta \omega_A/(2\nu)$, related to finite collisionality driving resistive DWs (RDWs), $\beta_\mu = \beta/(2\mu)$, associated with finite electron mass $m_e = \mu m_i$ driving inertial DWs (IDWs), and $\beta_A = (\omega_*/\omega_A)^2/(k_y\rho_s)^2$, being the ratio between the electron diamagnetic and the Alfvén frequencies. Linearly, DWs are strongly suppressed by electromagnetic effects with increasing $\beta_\nu$ (RDW branch) or $\beta_\mu$ (IDW branch)~\cite{Mosetto2012}. The parameter $\beta_A$, instead, describes the coupling between Alfvén waves and DWs, which becomes important for $\beta_A \sim 1$~\cite{Guzdar2001Alfven}. Balancing the inertial and resistive terms in Eq.~(\ref{eq:vpare_main_DW})~\cite{Ricci2010}, one finds that resistive effects dominate over inertial effects when $\beta_\nu /\beta_\mu = \mu \omega_A/\nu  \equiv \widehat{\mu} \ll 1$. Since $\widehat{\mu} \propto (T_e/n)^{3/2}$, the RDW-dominated regime, $\widehat{\mu} \ll 1$, corresponds to high density, as in the \texttt{GBS} simulations presented here, and the IDW-dominated regime, $\widehat{\mu} \gg 1$, to low density. Within the approximations used to derive Eqs.~(\ref{eq:density_main_DW}--\ref{eq:vpare_main_DW}), the evolution equation of the electric field, $\partial_t \langle \mathcal{E}_r \rangle = \mathcal{F}$, becomes
\begin{align}
    &\frac{\rho_s}{\sqrt{\beta_A} L_n}\frac{\partial}{\partial t} \frac{d \langle \phi \rangle}{d x}  + \sigma_T \frac{d \Pi}{ d x} = 0, \label{eq:zonal_flow_main_DW} 
\end{align}
where $L_n = 1/|d_x \log \langle n \rangle|$ and the turbulent stress, $\Pi \equiv \sum_{k_y} \rho_s^2 k_y^2 \,\Pi_{k_y}$, is given by
\begin{align}\label{eq:turb_stress}
    \Pi_{k_y} &= \left\langle  \frac{g}{\langle g \rangle}\hat s\left( \theta + \frac{r_0}{\hat s}\partial_x \theta - \sigma_T \Omega_E t\right) (|\phi|^2 - |A_\parallel|^2)\right\rangle ,
\end{align}
and is computed quasilinearly, by using the solutions for $\phi$ and $A_\parallel$ obtained from the linear problem, Eqs.~(\ref{eq:density_main_DW}--\ref{eq:vpare_main_DW}). In the chosen coordinates, $\langle f \rangle = \int_{-\tau/2}^{\tau/2} dt \int_{-\infty}^{\infty} d \theta f(\theta, t) / (\tau L_\theta)$, where $L_\theta = \int_{|f| > 0} d \theta$ and $\tau > 0$ is a time-integration interval, set by the characteristic lifetime of turbulence and chosen to be the decorrelation time due to flow shear, $\tau = 1/|\Omega_E|$~\cite{Biglari1990}. We note that neglecting the coupling to sound waves in Eqs.~(\ref{eq:density_main_DW}--\ref{eq:zonal_flow_main_DW}) removes the geodesic-acoustic-mode branch and isolates the quasi-stationary mean flow of interest.
The net momentum flux across the pedestal region $r \in (r_-, r_+)$ around the electric field minimum $r_{\rm min}$ ($r_\pm = r_{\rm min} \pm 2 \rho_s$) is obtained by integrating Eq.~(\ref{eq:zonal_flow_main_DW}), yielding $-\sigma_T [\Pi(r_+) - \Pi(r_-)]$. Mean flow growth, $\partial_t [\langle \phi \rangle(r_+) - \langle \phi \rangle(r_-)] > 0$, therefore requires $-\sigma_T [\Pi(r_+) - \Pi(r_-)] > 0$. In the RDW regime ($\widehat{\mu} \ll 1$), this condition is met once $\beta_\nu$ exceeds a critical value $\beta_\nu^*(\sigma_T, \beta_A, \Omega_E)$, determined numerically from the quasilinear model. It is found that $\beta_\nu^*$ depends only weakly on $\beta_A$ and $\Omega_E$: over the experimentally relevant range $\beta_A \in (1,40)$, $\Omega_E \in (0.1, 1)$, it varies by $\simeq 15\%$ about $\beta_\nu^*(\sigma_T=-1) \simeq 0.07$. The asymmetry ratio is found to be $A_\nu \equiv \beta_\nu^*(\sigma_T=+1)/\beta_\nu^*(\sigma_T=-1) \sim 2$ over the same range. Once the threshold $\beta_\nu > \beta_\nu^*$ is reached, the resulting rise in $\bm{E} \times \bm{B}$ shear acts to increase $\beta_\nu$ by suppressing turbulence, further increasing $\bm{E} \times \bm{B}$ shear. The quasilinear estimates agree well with the simulation results in Fig.~\ref{fig:fav_unfav_stresses_time}b, in which $\beta_\nu$ is shown at $r=r_{\rm min}$ as a function of time. The favourable configuration transitions to a suppressed transport regime for $\beta_\nu \simeq 0.09$ (at $t \simeq 10 \, R/c_s$), while the unfavourable configuration at matched power does not, transitioning instead at higher power, once $\beta_\nu \simeq 0.13$. In the low-density IDW-dominated regime ($\widehat{\mu} \gg 1$), though not simulated here with \texttt{GBS}, a transition is predicted quasilinearly to occur once $\beta_\mu > \beta_\mu^*(\beta_A, \Omega_E)$, with $\beta_\mu^*$ again weakly dependent on $\beta_A$ and $\Omega_E$, varying by $\simeq 10\%$ about $\beta_\mu^* \simeq 0.06$.

We now discuss the symmetry properties of Eqs.~(\ref{eq:density_main_DW}--\ref{eq:zonal_flow_main_DW}). The transformation $\sigma_p \to -\sigma_p, A_\parallel \to -A_\parallel$ leaves the system invariant, implying that $\Pi^{\sigma_p} = \Pi^{-\sigma_p}$ and reproducing the experimental observation that the direction of the plasma current does not influence the L--H power threshold~\cite{Takizuka2004}. Indeed, we verify that reversing the sign of $\sigma_p$ does not affect the \texttt{GBS} simulation results. Meanwhile, the field-reversal symmetry $\sigma_T \to -\sigma_T$ is broken only when four conditions hold simultaneously: (i) up–down geometric asymmetry, $g(\theta) \neq g(-\theta)$, which breaks the $(\sigma_T, \sigma_p, k_y, \theta) \to (-\sigma_T, -\sigma_p, -k_y, -\theta)$ symmetry mapping $\Pi \to -\Pi$; (ii) finite $\bm{E} \times \bm{B}$ shear ($\Omega_E \neq 0$) and (iii) finite magnetic shear ($\hat s \neq 0$), which together break the  $(\sigma_T, k_y) \to (-\sigma_T, -k_y)$ symmetry; and (iv) finite collisionality ($\nu > 0$), which breaks the time-reversal invariance $(\sigma_T, \sigma_p, t) \to (-\sigma_T, -\sigma_p, -t)$. This result generalises the time-independent Reynolds-stress analyses of~\cite{Fedorczak2012, Grover2024, Frei2026}, which identify (i)--(iii) but treat $k_\perp^2$ in Eq.~(\ref{eq:kperp}) as constant in time, thereby missing condition (iv).

Using the transition criteria derived from the quasilinear theory, $\beta_\nu > \beta_\nu^*(\sigma_T)$ and $\beta_\mu > \beta_\mu^*$, we now obtain analytical scaling laws for the L--H power threshold in the high- and low-density branches, as well as the density minimum and its associated power. Indeed, since $\beta_\nu \sim T_e^{5/2}/\sqrt{n}$ and $\beta_\mu \sim n T_e$, a transition requires the edge temperature to exceed a critical value that increases with $n$ in the RDW branch and decreases with $n$ in the IDW branch; through the power balance derived below, $P_{\rm LH}$ thus increases with density in the former and decreases in the latter, producing the density minimum. 

The power crossing the last-closed-flux-surface is given by $P = \int Q_\perp d S$, with $Q_\perp$ the radial heat flux density, integrated over the surface $S$. Using a mixing-length estimate for turbulent transport, we have $Q_\perp \sim \gamma |\partial_x p|/k_x^2 \sim \gamma p / k_y$,
where the nonlocal linear scaling relation $k_x \sim \sqrt{k_y/L_p}$ is employed~\cite{Rogers2005, Ricci2008, Tecchiolli2025}, valid for both DW and ballooning instabilities. Substituting the expression of $Q_\perp$ into $P \sim Q_\perp S$ yields, for DW turbulence with $\gamma/k_y \sim \rho_s c_s/L_n$~\cite{Ricci2010}, $P \sim   S \rho_s c_s  n T_e /L_n$. To express the edge temperature as a function of input power, we note that the energy crossing the separatrix is removed from the system via parallel losses, $P \sim 2\pi R \int_\Delta Q_\| \bm{b} \cdot \bm{B}_p/|B_p|dl$, where $\Delta \sim L_p$ is the scrape-off-layer width, $\bm{B}_p$ is the poloidal magnetic field and $Q_\parallel$ is the parallel heat flux density~\cite{Fundamenski2012}. Given that $\bm{b} \cdot \bm{B}_p/|B_p| \sim a/(q R)$, we find $P \sim 2\pi a Q_\|\Delta/q$. Depending on the plasma collisionality at the separatrix, $Q_\parallel$ is either dominated by conductive transport or by sheath losses~\cite{Stangeby2000}. In the conduction regime (typical operation regime in tokamaks~\cite{Fundamenski2012}), $Q_\| \sim \chi_{\|e}\nabla_\|T_e \sim \chi_{\|e}  T_e / L_\|$, where $\chi_{\|e} = 0.95(4\pi\epsilon_0/e^2)^2 T_e^{5/2}/(\sqrt{m_e}Z_{\rm eff}\Lambda) \equiv \chi_0 T_e^{5/2}$ is the parallel electron heat diffusivity and $\Lambda$ is the Coulomb logarithm. One obtains in this way a relation between the power and the gradient length-scales, $L_n \sim L_p$, given by $P \sim n S  \left[7qL_\|P/(4\pi a L_p\chi_0)\right]^{4/7}/ (e B L_p)$, which can be used to express $\beta_\nu$ and $\beta_\mu$ as functions of engineering variables only, namely $m_i, n, B, a, R, q, Z_{\mathrm{eff}}$ and $P$.
\begin{figure}[htbp]
  \centering
  \includegraphics[width=0.49\columnwidth]{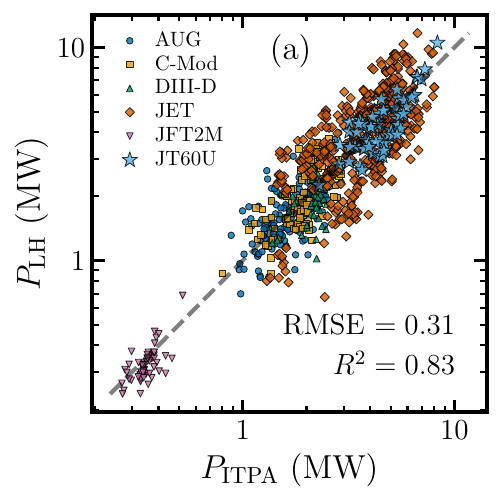}%
  \hfill
  \includegraphics[width=0.49\columnwidth]{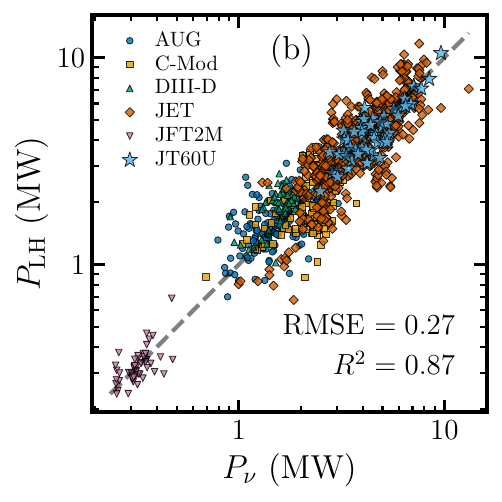}%
  \\
  \includegraphics[width=0.49\columnwidth]{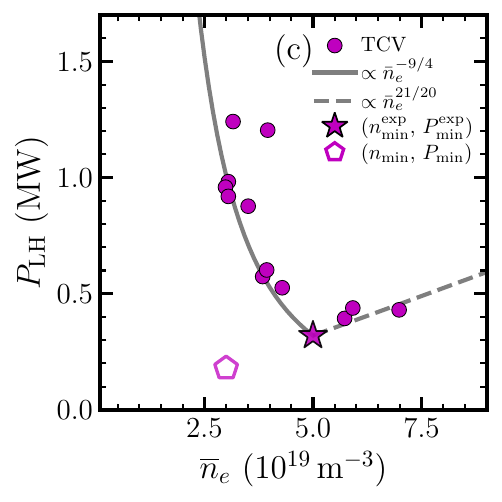}%
  \hfill
  \includegraphics[width=0.49\columnwidth]{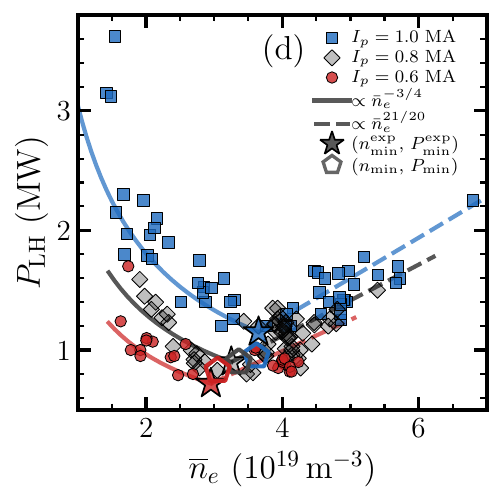}%
  \\
  \includegraphics[width=0.49\columnwidth]{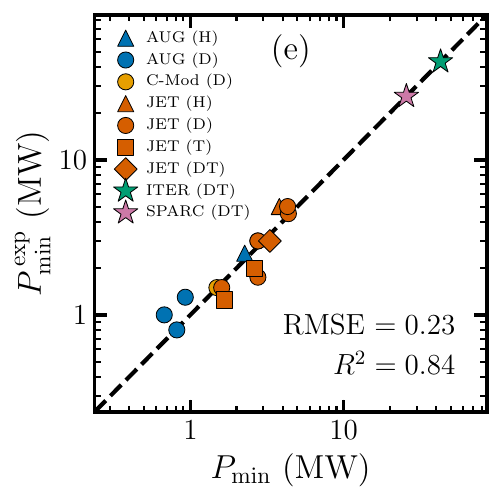}%
  \hfill
  \includegraphics[width=0.49\columnwidth]{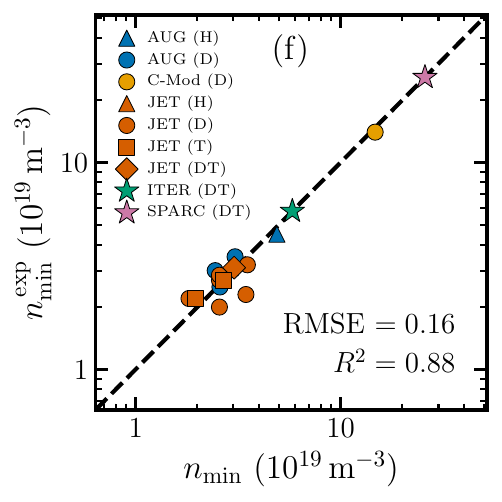}
  
  \caption{Favourable configuration L--H threshold:  (a) empirical from Eq.~(\ref{eq:ITPA}), (b) analytical from Eq.~(\ref{eq:Pd}) (using $\Lambda = 15$ and, when missing, $Z_{\rm eff} = 1.5$). Low- and high-density branches in TCV (c) and AUG (d) (data from~\cite{Piras2010, Ryter2013}). $P_{\mathrm{min}}$ (e) and $n_{\rm min}$ (f) (data from~\cite{Delabie_2026}).}
\label{fig:scaling}
\end{figure}
 By imposing the threshold criteria $\beta_\nu \gtrsim \beta_\nu^*(\sigma_T)$ (RDW), and $\beta_\mu \gtrsim \beta_\mu^*$ (IDW), one obtains expressions for the L--H power in the high-density branch,
\begin{align}
P_\nu &= K_\nu \frac{\beta_\nu^*(\sigma_T)^{11/10} e^{19/10}\Lambda^{3/5}m_e^{3/10}}{\mu_0^{11/20}\epsilon_0^{6/5}m_h^{11/20}} \times   \nonumber \\
&\qquad \qquad \frac{a R^{11/10} B^{3/5}n^{21/20} Z_{\rm eff}^{3/5} q^{1/10}}{M^{11/20}}, \label{eq:Pd}
\end{align}
and in the low-density branch,
\begin{align}
P_\mu &= K_\mu  \frac{\beta_\mu^{* 11/4} \epsilon_0 m_e^{5/2}}{e^{5/2}\sqrt{\Lambda}\mu_0^{11/4}m_h^{11/4}} \frac{aB^5 }{n^{9/4}q\sqrt{Z_{\rm eff}}M^{11/4}}, \label{eq:PI}
\end{align}
respectively. In Eqs.~(\ref{eq:Pd}) and~(\ref{eq:PI}), the surface area and connection length are estimated as $S \simeq 4\pi^2 a R$ and $L_\parallel \simeq 2 \pi q R$, the main ion mass number is $M = m_i/m_h$, with $m_h$ the hydrogen mass, and the dimensionless factors $K_\nu$ and $K_\mu$ account for $O(1)$ scaling approximations in the derivation. While all exponents in Eqs.~(\ref{eq:Pd}) and~(\ref{eq:PI}) are derived without regression against threshold data, these constants are free parameters and are fitted to experiments. In Fig.~\ref{fig:scaling}b, Eq.~(\ref{eq:Pd}) is compared with the ITPA database of high-density, favourable-configuration transitions~\cite{Martin2008}. With the quasilinear $\beta_\nu^*(\sigma_T=-1) = 0.07$ and the fitted $K_\nu = 0.325$, we obtain $R^2 = 0.87$. The empirical scaling law, Eq.~(\ref{eq:ITPA}), achieves instead $R^2 = 0.83$ (Fig.~\ref{fig:scaling}a). We note that the theoretical scaling law in Eq.~(\ref{eq:Pd}), $P_\nu \sim a \, R^{11/10} B^{3/5} n^{21/20}$, also agrees with the recently revised ITPA scaling law for metal-wall devices (ITER-like-wall), found to scale as $P^{\rm ITPA}_{\rm ILW} \sim a \, R \, B^{0.58} n^{1.08}$~\cite{Delabie_2026}. The effective charge exponent $Z_{\rm eff}^{3/5}$ in Eq.~(\ref{eq:Pd}) also agrees with the experimentally observed dependence $\sim Z_{\rm eff}^{0.7}$~\cite{Takizuka2004,Birkenmeier2026}. Combining Eq.~(\ref{eq:Pd}) with the quasilinear results for the asymmetry ratio, $A_\nu \sim 2$, the power asymmetry factor is approximately $P_\nu(\sigma_T=+1)/P_\nu(\sigma_T=-1) = A_\nu^{11/10} \sim 2$, consistent with experimental observations~\cite{Takizuka2004}. 

The high- and low-density threshold expressions, Eqs.~(\ref{eq:Pd}) and~(\ref{eq:PI}), also agree well with TCV data (Fig.~\ref{fig:scaling}c). At higher power, the low-density branch can enter a sheath-limited rather than conduction-limited regime ($Q_\parallel \sim n T_e c_s$), reached when $P[\mathrm{MW}]/(n[10^{19}\,\mathrm{m}^{-3}])^{7/4} \gg 8\pi a L_p L_\parallel^{3/4}/q$ (lengths in m)~\cite{Stangeby2000}. In this regime, the low-density scaling is modified to $P_\mu^\prime = K_\mu^\prime (m_e^{7/4}/\sqrt{e}\, m_h^2 \mu_0^{7/4})a B^3 R^{1/2} n^{-3/4} Z^{1/4} M^{-2}$, with a weaker density dependence than Eq.~(\ref{eq:PI}), consistent with observations in AUG and JET~\cite{Ryter2013, Maggi2014}. In AUG, $8\pi a L_p L_\parallel^{3/4}/q \simeq 1$, thus for typical low-density scenarios~\cite{Eich2021, Cavedon2025} with $n_{\rm sep} \lesssim 10^{19}\,\mathrm{m}^{-3}$, the sheath-limited regime occurs for $P \gtrsim 1\,\mathrm{MW}$. Indeed, $P_\mu^\prime$ describes AUG data~\cite{Ryter2013} across three values of $I_p$ remarkably well (Fig.~\ref{fig:scaling}d). Though kinetic effects are expected to become important at low enough density, we note that $\lambda_e/L_\parallel > 1$ is not equivalent to $\widehat\mu  > 1$: for $\beta_\mu = \beta_\mu^*$, $\widehat{\mu} \simeq 20 \pi (\lambda_e/L_\parallel)$, so a collisional edge plasma with $\lambda_e/L_\parallel \simeq 0.1$ can still lie in an IDW-dominated regime, $\widehat{\mu} \simeq 6$. 

The low- and high-density branches of the L--H transition coincide when inertial and resistive effects are comparable, $\widehat{\mu} \sim 1$. This yields a scaling law for the density minimum $n_{\rm min}$,
\begin{align}\label{eq:density_min}
    n_{\rm min} = K_n  \frac{\epsilon_0^{2/3} m_e^{2/3}}{e^{4/3} \mu_0^{1/3} \Lambda^{1/3} m_h^{2/3}}\frac{I_p^{1/3} B}{Z_{\rm eff}^{1/3}  a^{2/3} M^{2/3}},
\end{align}
where the power is evaluated with the high-density scaling law, Eq.~(\ref{eq:Pd}). Substituting Eq.~(\ref{eq:density_min}) in Eq.~(\ref{eq:Pd}), one finds an expression for the minimum power required to access an H--mode regime,
\begin{align}\label{eq:PLH_min}
    P_{\rm min} = K_P\frac{e^{1/2}  \Lambda^{1/4} m_e}{\epsilon_0^{1/2}   m_h^{5/4} \mu_0}\frac{a^{1/2} R \, B^{7/4} I_p^{1/4} Z_{\rm eff}^{1/4} }{ M^{5/4}},
\end{align}
where the safety factor is expressed in terms of the plasma current $I_p$, via $q \sim 2 \pi a^2 B/(\mu_0 R I_p)$. Eqs.~(\ref{eq:density_min}) and~(\ref{eq:PLH_min}) are consistent with existing empirical scalings~\cite{Ryter2014}; moreover, fitting $K_n = 1.21$, which fixes $K_P = K_\nu \beta_\nu^{*}(\sigma_T=-1)^{11/10} K_n^{21/20} = 0.0213$, they describe the favourable-configuration data for ITER-like-wall devices (Table~1 of~\cite{Delabie_2026}) very well (see Figs.~\ref{fig:scaling}e and~\ref{fig:scaling}f). Since the favourable/unfavourable symmetry is broken only by the resistive term in Eq.~(\ref{eq:vpare_main_DW}), $\beta_\mu^*$ is independent of $\sigma_T$ in the IDW limit. However, Eqs.~(\ref{eq:Pd}) and~(\ref{eq:PI}) are derived in the asymptotic limits $\widehat{\mu} \ll 1$ and $\widehat{\mu} \gg 1$: around the density minimum, $\widehat{\mu} \sim 1$, resistive and inertial effects remain comparable, and the asymmetry is expected to survive, as observed in AUG~\cite{Plank2023}. 

In TCV, the measured $P_{\rm min}^{\rm exp}$ and the line-averaged $n_{\rm min}^{\rm exp}$ exceed the predicted values (Fig.~\ref{fig:scaling}c). The power offset is plausibly due to unsubtracted radiative losses~\cite{Piras2010}. For the density, our theory predicts the edge value, which in~\cite{Piras2010} is about half of $\overline{n}_e$; the plotted line-averaged $n_{\rm min}^{\rm exp}$ is thus twice the predicted edge $n_{\rm min}$ (in AUG instead, $n_{e\,95} \simeq \overline{n}_e$~\cite{Ryter2013}). The lower $P_{\mathrm{LH}}$ found for NBI-heated transitions in the low-density branch~\cite{Ryter2014} is also captured: reduced electron heating lowers the edge $T_e$ and shifts the $\widehat{\mu} \sim 1$ boundary to lower density, decreasing $P_{\mathrm{LH}}$ at fixed $n$. We note that reference~\cite{Ryter2014} instead places the density minimum where the ion-to-electron collisional time reaches a fixed fraction of the confinement time, later interpreted~\cite{Malkov2015, Bilato_2020} as the point where a critical edge shear ($E_r \sim \partial_r p_i/(Z e n)$) is reached at the lowest input power. The mechanism proposed here (sudden mean-flow generation by finite-$\beta$ DW turbulence, closer to the picture of~\cite{Diamond1994, Rogers1998, Guzdar2001}) reproduces the same data and further explains why $E_r$ does not seem to increase prior to the L--H transition in JET and NSTX~\cite{Silva2021, Zweben2021}. Finally, although the isotope scaling of the threshold power, $P_{\nu} \sim M^{-11/20}$ in Eq.~(\ref{eq:Pd}), is weaker than the observed trend, $P_{\mathrm{LH}}\sim 1/M$~\cite{Delabie_2026}, the minimum power $P_{\rm min} \sim M^{-5/4}$ in Eq.~(\ref{eq:PLH_min}) closely matches the empirical exponent. 

Extrapolating Eqs.~(\ref{eq:density_min}) and~(\ref{eq:PLH_min}), calibrated on present devices, to ITER's full-field scenario ($I_p = 15$ MA, $B = 5.3$ T) with $Z_{\rm eff} = 1.5$ and $M=2.5$, we obtain $n_{\rm min} = (5.8 \pm 1.0) \times 10^{19}~\mathrm{m}^{-3}$ and $P_{\rm min}  = (43 \pm 11)$ MW, with uncertainties given by the root-mean-square errors of the scalings. ITER should thus access H--mode within its $73$ MW auxiliary heating limit~\cite{Loarte_2025}. In the full-field scenario of SPARC ($I_p = 8.7$ MA, $B = 12.2~\mathrm{T}$)~\cite{Creely_2020}, we find instead $n_{\rm min} = (2.6 \pm 0.4) \times 10^{20} \, \mathrm{m}^{-3}$ and $ P_{\rm min}  = (26 \pm 7) \, \mathrm{MW}$, leaving little margin over its planned $25 \, \mathrm{MW}$ heating capacity, a concern raised in Ref.~\cite{Hughes2025} from empirical scalings and rederived here from a physics-based theory. 

\textit{Acknowledgments.} We thank F. Ryter for providing access to AUG data and for his insightful comments, as well as L. Aucone, A. Balestri, J. Ball, S. Brunner, B. Frei, M. Giacomin, T. Görler, S. García Herreros, Y. Martin, M. Pedrini, E. Tonello, and the TSVV-A group for helpful discussions. The simulations in this work were run on Daint (s$1170$), Daint-ALPS (lp$129$) and Discoverer (EHPC-REG-$2023$-R$01$-$138$). This work has been carried out within the framework of the EUROfusion Consortium, partially funded by the European Union via the Euratom Research and Training Programme (Grant Agreement No 101052200 — EUROfusion).
The Swiss contribution to this work has been funded in part by the Swiss State Secretariat for Education, Research and Innovation (SERI).
Views and opinions expressed are however those of the author(s) only and do not necessarily reflect those of the European Union, the European Commission or SERI. 
Neither the European Union nor the European Commission nor SERI can be held responsible for them. This work was supported in part by the Swiss National Science Foundation.

%

\end{document}